%% file: paper.tex
\documentclass[12pt,article,aps, amsfonts, nofootinbib, notitlepage]{revtex4-2}
\input{mylatex.tex}

\begin{document}

\title{Kolmogorov-Arnold Wavefunctions}

\author{Paulo F. Bedaque}
\email{bedaque@umd.edu}
\affiliation{Department of Physics,
University of Maryland, College Park, MD 20742}

\author{Jacob Cigliano}
\email{cigliano@terpmail.umd.edu}
\affiliation{Department of Physics,
University of Maryland, College Park, MD 20742}

\author{Hersh Kumar}
\email{hekumar@umd.edu}
\affiliation{Department of Physics,
University of Maryland, College Park, MD 20742}

\author{Srijit Paul}
\email{spaul137@umd.edu}
\affiliation{Department of Physics,
University of Maryland, College Park, MD 20742}

\author{Suryansh Rajawat}
\email{suryansh@terpmail.umd.edu}
\affiliation{Department of Physics,
University of Maryland, College Park, MD 20742}

\preprint{}
\begin{abstract}

This work investigates Kolmogorov-Arnold network-based wavefunction ansatz as viable representations for quantum Monte Carlo simulations. Through systematic analysis of one-dimensional model systems, we evaluate their computational efficiency and representational power against established methods. Our numerical experiments suggest some efficient 
training methods and we explore
 how the computational cost scales with desired precision, particle number, and system parameters. 
Roughly speaking, KANs seem to be 10 times cheaper computationally than other neural network based ansatz.
 We also introduce a novel approach for handling strong short-range potentials—a persistent challenge for many numerical techniques—which generalizes efficiently to higher-dimensional, physically relevant systems with short-ranged strong potentials common in atomic and nuclear physics.

\end{abstract}
\maketitle

\section{Introduction}
Recent advances in machine learning (ML) have opened new avenues for solving complex problems in quantum many-body physics. A particularly promising direction lies in the interplay between neural networks and variational quantum Monte Carlo (VMC) methods. By leveraging flexible neural network architectures, one can construct highly general ansätze for many-body wave functions, capable of capturing intricate correlations without relying on restrictive physical assumptions, in a computationally efficient way.

This connection arises from a deep analogy between VMC and unsupervised machine learning: in both frameworks, an optimization process minimizes a target function—whether it be the energy of a quantum system (in VMC) or a cost function (in ML)—by iteratively adjusting model parameters. In VMC, the energy serves as the cost function, and the neural network parametrizes the trial wave function, with the variational minimum providing an upper bound to the true ground-state energy. The stochastic nature of VMC, where energies and gradients are estimated via Monte Carlo sampling, further mirrors the optimization dynamics of large-scale ML training. This analogy can be pushed further still by noticing that, in many ML tasks, a regularizing term is added to the cost function in order to avoid overfitting. This term, which disfavors highly twisted functions, can take a form identical to the kinetic term in quantum mechanics. Seen this way, the problem of finding quantum mechanical ground states is identical to the one of minimizing the potential energy with a regularization term that smooths out the wavefunction by an amount controlled by the value of $\hbar$.

The idea of parametrizing ground state wavefunctions using some form of neural network has been explored extensively lately, starting with Restricted Boltzmann Machines representing ground states of spin systems with discrete variables  \cite{carleo}. Systems of particles moving in space, with their continuous coordinates have also been comprehensively explored, in particular in the case of long-range (Coulomb) forces common in condensed matter and quantum chemistry (for instance, \cite{paulinet,ferminet, Nys}). The attention to systems with short-range forces  is a little more recent \cite{lovato1,lovato2,lovato3,lovato4,lovato5,lovato6,Wen:2025mlq,Wang:2024ynn,Keeble:2023rre,Yang:2022rlw,Keeble:2019bkv,fermionpaper} and mostly motivated by nuclear applications.

More recently, a new kind of  neural network architecture, Kolmogorov-Arnold Networks (KAN) was put forward \cite{KAN-MIT} as presenting several advantages over the more common feed-forward/multilayer perceptron (MLP) architecture 
(for a review see \cite{hou2024comprehensivesurveykolmogorovarnold}). 
This architecture, which theoretically enables efficient representation of multivariate functions, has yet to be  explored in the context of quantum many-body problems. Testing such models within VMC could yield more expressive wave functions while maintaining computational tractability—a crucial advantage given the high-dimensional parameter spaces often encountered in many-body physics. By integrating modern ML techniques, VMC can be scaled efficiently, enabling accurate simulations of large systems even with modest computational resources. The main goal of this paper is to explore the use of KANs in variational quantum mechanical calculations. We will do this in the context of simple models involving many bosons moving in a dimensional trap and interacting with short-ranged and long-range forces. Some of these models are soluble and provide a solid test of our methods; others are not soluble but are closer to actual physical systems found in cold atomic traps and are a natural stepping stone towards realistic nuclear models. In \sect{sec:KAN-MLP} we review the architectures of our ansatz and discuss issues related to training. We discuss the pros and cons of KAN's vs. MLP's parametrizations in \sect{sec:comparison}. A problem related to cusps in the wavefunction, which is just a more severe issue encountered with all strong short-range potentials, is bypassed in \sect{sec:2body}. A summary of the results and a discussion of future work is presented in \sect{sec:conclusion}. An important technical detail regarding the sampling of $\delta$-function potentials is discussed in the Appendix.

\section{Kolmogorov-Arnold and Feed-Forward Networks}\label{sec:KAN-MLP}

\begin{figure}
    \centering
    \includegraphics[width=0.6\linewidth]{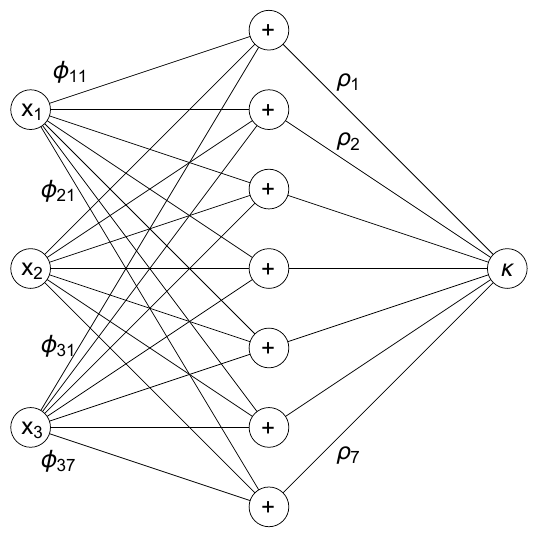}
    \caption{Kolmogorov-Arnold network with 3 inputs and $2 \times 3+1=7$ hidden nodes. The name of some of the one-variable functions are shown on the corresponding links. The bosonic KAN is the special case where $\phi_{11}=\phi_{21}=\cdots=\phi_{71}$, $\phi_{12}=\phi_{22}=\cdots=\phi_{72}$ and so on.\textcolor{red}
    }
    \label{fig:KAN}
\end{figure}

Kolmogorov-Arnold networks (KANs) are a class of neural networks inspired by the Kolmogorov-Arnold representation theorem, which states that any multivariate continuous function can be represented as a {\it finite} composition of univariate functions and additions.
More precisely \cite{kolmogorov,arnold,hou2024comprehensivesurveykolmogorovarnold}, the theorem states that any continuous function of $N$ variables $\kappa(x_1, \cdots, x_N)$ can be represented as
\beq \label{eq:KAN}
\kappa(x_1,...x_N) = \sum^{2N+1}_q \rho_q\left(\sum^N_p \phi_{qp}(x_p)\right),
\eeq where $\rho_q(x)$ and $\phi_{pq}(x)$ are one-variable continuous functions. Some versions of the theorem even allow for the functions $\phi_{pq}(x)$ to be fixed while only $\rho_q(x)$ is varied to represent different $\kappa(x_1,...x_N)$. The composition of functions in \eq{eq:KAN} is better visualized by a diagram, such as the one in \fig{fig:KAN}, where the functions $\rho_q(x)$ and $\phi_{pq}(x)$ are represented by {\it lines} while the nodes represent  the addition of all its inputs.  For this reason we will call the functions $\rho_q(x), \phi_{pq}(x)$ (and their generalizations) the ``line-functions". This graphical representation is to be contrasted with the depiction of more common multi-layer perceptron/feed-forward networks where the lines represent numerical weights and nodes represent fixed  activation functions.

The Kolmogorov-Arnold theorem guarantees that $\rho_q(x)$ and $\phi_{pq}(x)$ are continuous but, in many cases, they seem to have a complicated fractal-like structure. This poses serious doubt about their use in machine learning tasks, a point extensively discussed in the literature (see, for instance 
\cite{10.1162/neco.1989.1.4.465,kurkova}).
It is  unknown to us whether imposing further regularity constraints on $\kappa(x_1, \cdots, x_N)$ also constrains the line-functions to be smoother. We do show below evidence that this is the case, at least in an approximate sense.
The applicability of the Kolmogorov-Arnold theorem to numerical tasks was revived recently by the demonstration that KANs with more than one hidden layer and smooth line-functions can be profitably used in several  artificial intelligence problems, showing improved scaling and interpretability \cite{MIT-KAN}. 

The practical use of KANs requires the line-functions  to be parameterized. We follow \cite{MIT-KAN} and approximate them by splines (see \cite{deboor1978practical, dierckspline} for a quick introduction).
A spline of degree $q$ is defined as a piecewise function on a given interval, where each piece is a polynomial of degree $q$. We choose to use quadratic splines ($q=2$) to guarantee the regularity expected of wavefunctions (more about this below). The parameters uniquely\footnote{Actually, there is one degree of freedom granted by continuity and differentiability requirements. A common method to remove this extra freedom, and one we adopt, is demanding the first polynomial to be linear. See Reference \cite{deboor1978practical}} specifying a spline are their values at a set of equidistant points called knots. The different polynomials on each interval are chosen such that they pass through a chosen value at each knot, and such that the total piecewise function is continuously differentiable ($q-1$) times. The set of all chosen values serves as a parametrization for a full KAN.

In order to connect different layers we need the image of one layer to lie in the domain of the next. We choose the domain of all the line functions to be
 $(-1,1)$ and use the expression
 \beq \label{eq:KAN2}
 \kappa(x_1, ..., x_N) = \sum^{2N+1}_q \alpha_q\left(\tanh \left( \frac{1}{N} \sum^N_p \beta_{qp}\left(\tanh(x_p)\right)\right)\right)
=
\sum^{2N+1}_q \rho_q\left(\sum^N_p \phi_{qp}(x_p)\right)
\eeq with $\rho=\alpha\circ \tanh\circ\frac{1}{N}$ and $\phi=\beta \circ\tanh$.
 The insertion of 
$\tanh(x)$ above guarantees that the output of the first layer lies within the domain of the next spline. A factor of $1/N$ is also included to ensure the input to $\tanh(x)$ is of $\mathcal{O}(1)$. This factor guarantees that $\tanh(x)$ varies sufficiently across the region where the wavefunction has significant amplitude, to avoid suppressing any key behaviors. \eq{eq:KAN2} and the associated spline parameters are the KAN analog of a neural network architecture and its weights and biases.

KAN's have a particularly convenient property, which we will exploit: they can be scaled in an obvious and direct manner, as follows. New knots are inserted between existing ones. The value at each new knot is chosen to equal the current value of the spline at that point. Each resulting finer spline uniquely satisfies all given requirements (see above) and agrees everywhere with the coarser previous spline. This procedure grants the KAN more expressibility and permits significant computational benefit, as a fine KAN is only required during training once a coarse KAN has plateaued. Continuity and stability during training are guaranteed by equality of the coarser and finer KANS, with the finer KAN inheriting the features of its predecessor, without requiring retraining from scratch.

For the purpose of comparison, we will also use multi-layer perceptrons/feed-forward networks.
A MLP with $L$ layers and which takes $n_1$ inputs and produces $n_L$ outputs is defined as 

\beq
I_{i+1} = O_i = \sigma(W_i \Vec{I}_i + \Vec{b}_i), \;  1 \leq i \leq L-1,
\quad
O_L = W_L\Vec{I}_L + \Vec{b}_L.
\eeq
$I_i$ are the inputs for layer $i$ that are multiplied by a weight $n_{i+1} \times n_i$ matrix $W_i$ and then added to a bias vector $b_i$ of length $n_{i+1}$. This transformation, composed with a fixed activation function $\sigma$, results in output $O_i$ which serves as the input for the ($i+1^{th}$) layer. The output from the final layer has no activation function.

The wave function for a system of $N$ indistinguishable bosons is symmetric under particle exchanges: 
\beq\psi(...,x_i,...,x_j,...) = \psi(...,x_j,...,x_i,...).
\eeq Exchange symmetry can be enforced easily on a KAN by setting the line-function of the first layer ($\alpha_{pq}$) to be independent of $q$:
\beq
\kappa(x_1, ..., x_N) = \sum^{2N+1}_q \rho_q\left(\sum^N_p \phi_{q}(x_p)\right)
\eeq
The other layers, if present, are unrestricted. We call a network of this kind a bosonic KAN.
This results in a  totally symmetric function $\kappa(x_1,\cdots, x_N)$  depending on  less parameters than the generic one KAN. 
We are unaware of a
 proof of the reasonable assumption that a bosonic KAN is an universal approximator of any  {\it symmetric} function.
 For a MLP the symmetrization can be done, in one dimension, by ordering the coordinates, as we do in this paper using the method described in \cite{neuralnetworkpaper} or using more involved methods as in \cite{Corey}.

\section{Numerical results and comparison}\label{sec:comparison}

The first model we consider is given by the Hamiltonian:
\beq\label{eq:astra}
H = \sum_i^N \left(
-\frac{1}{2m} \frac{\partial^2}{\partial x_i^2} + \frac{1}{2}m\omega^2x_i^2
\right) + 
\sum_{i<j} \left(g \delta(x_i - x_j) + \sigma |x_i - x_j| \right).
\eeq
describing $N$ bosons in a one-dimensional harmonic trap with both long-range and
a  short-range  interactions.  The usefulness of this model comes from the fact that it is soluble in the $\sigma = -m\omega g/2$ case \cite{exactboson1d}. The energy and wave functions of the ground state in that case\footnote{Reference \cite{exactboson1d} claims the model is soluble when $\sigma=-m\omega g$ but we find analytically that the provided solution is correct for $\sigma=-m\omega g/2$} are given by: 
\bea\label{eq:astra-exact}
E_0 &=& \frac{N\omega}{2} - mg^2\frac{N(N^2 -1)}{24}, \\
\psi_0(x_1,\cdots,x_N) 
&=& 
 \prod_{i<j} e^{mg|x_i - x_j|/2} \prod_i e^{-m\omega x_i^2/2}.
\eea

Notably, the ground state energy does not depend on the sign of $g$, so we only consider the $g>0$ case from now on. Based on the exact values of the energy, we 
can distinguish a perturbative regime where $\frac{mg^2(N^2-1)}{12} \alt 1$ and a non-perturbative one where $\frac{mg^2(N^2-1)}{12} \agt  1$.

Our choice of ansatz is based on the universality of KANs: any real and positive function can be represented as $e^{-\kappa(x_1, \cdots, x_N)}$ and we assume that any real, positive {\it bosonic} wavefunction can be represented the same way by a {\it bosonic} KAN. By increasing the number of knots on the splines defining the line-functions in the KAN one can, in principle, reproduce any valid wavefunction. In practice, an extra adjustment make the ansatz somewhat easier to train. Based on the expected asymptotic behavior of the ground state  at large values $x_1^2+\cdots+x_n^2$, we write our final ansatz
as

\beq\label{eq:kan-ansatz} \psi(x_1,...,x_N) = e^{-\alpha(x_1^2 + \cdots +x_N^2)}e^{-\kappa(x_1, \cdots, x_N)},
\eeq
where $\kappa(x_1,...,x_N)$ is a bosonic KAN network as described earlier and $\alpha$ is an additional variational parameter.
The parameter $\alpha$ is treated as a variational parameter to be changed during training. Its main role is to make the wavefunction normalizable for a vast class of KAN $\kappa(x_1,...,x_N)$, which is required in the beginning of the training where the KAN is far from optimized.  The explicit gaussian term in \eq{eq:kan-ansatz}, although not strictly necessary, improves the convergence of the training process significantly.
We initialize the minimization of the energy with $\alpha=\frac{1}{2}\sqrt{m\omega}$ and the parameters of the KAN (i.e the spline parameters) are chosen randomly from a uniform distribution to be $\mathcal{O}(1)$. 

We vary the parameters of the ansatz, $\theta$, to minimize $\langle E \rangle$ using gradient descent methods, more specifically, the ADAM optimizer \cite{adam} with learning rates ranging from $10^{-3}$ to $10^{-7}$. The integrals for computing $\langle E \rangle$ and $\langle \partial E/\partial\theta \rangle$ are estimated using the Metropolis algorithm drawing particle positions $\Vec{x} = (x_1,...,x_N)$ from the probability distribution $|\psi_\theta(\Vec{x})|^2/\int d^N\Vec{x} \ |\psi_\theta(\Vec{x})|^2$.

As in many machine learning frameworks, the specific design of the training procedure plays a pivotal role in the success of variational approaches. Following extensive empirical testing, we have established a training protocol that balances model expressivity, statistical efficiency, and numerical stability throughout the optimization trajectory.

During the early stages of training, high-precision estimates of observables—particularly the energy and its gradient—are not essential. Instead, we rely on relatively coarse Monte Carlo statistics, which provide sufficient signal to guide the variational parameters toward improved regions of the energy landscape. Higher statistical accuracy is reserved for the later stages, where convergence to a highly accurate wavefunction demands finer resolution in the number of spline knots.

To facilitate convergence and enhance generalization across physical regimes, we adopt a staged training approach with respect to the coupling constant $g$. Beginning with a small value of $g$, we gradually increase it during training, akin to a continuation method. This strategy, similar in spirit to protocols used in Refs.~\cite{saito,neuralnetworkpaper,fermionpaper}, enables faster convergence to the target regime while yielding intermediate wavefunctions that can be efficiently fine-tuned for a range of $g$ values.

The training begins with a simplified model: a small value of $g$, a low-resolution variational ansatz (characterized by a limited number of spline knots), and modest Monte Carlo sampling. As the optimization proceeds and the wavefunction improves, we increase the sampling statistics and the coupling strength. At key points—identified by saturation in the training energy, indicating a bottleneck in expressivity—we enhance the variational flexibility by doubling the number of spline knots, using the scheme outlined in the previous section.


To maintain numerical stability, the learning rate is progressively reduced as the number of knots and samples increases. These changes must be carefully synchronized: increasing spline resolution without a corresponding increase in sampling fidelity can amplify statistical noise, leading to unreliable gradients and poor convergence. For example, since the first layer of splines directly processes particle coordinates, high resolution implies that even slight differences between configurations must be statistically distinguishable. Accurate gradient estimation under these conditions necessitates correspondingly high sampling precision.

In all cases presented in this work, the number of spline knots remains below $300$. Additional Monte Carlo parameters, including the proposal step size, thermalization length, and de-correlation skipping, are also tuned adaptively throughout training. The proposal step size is dynamically adjusted to maintain an acceptance rate near 50\%, consistent with established best practices in Variational Monte Carlo. These combined strategies yield a robust and efficient training procedure capable of producing high-fidelity variational wavefunctions across a range of physical couplings.
Since the variational parameters—and consequently the wavefunction and its associated probability distribution—are updated incrementally during training, the distribution changes only slightly between successive optimization steps. Therefore, extensive re-thermalization after each update is unnecessary. To enhance sampling efficiency, we initialize the Markov chain at each training step using the final configuration from the previous step. The chain is then allowed to undergo a short thermalization phase to adapt to the updated distribution before generating new samples for gradient estimation.

\begin{figure}[h]
    \centering
    \includegraphics[width=0.6\linewidth]{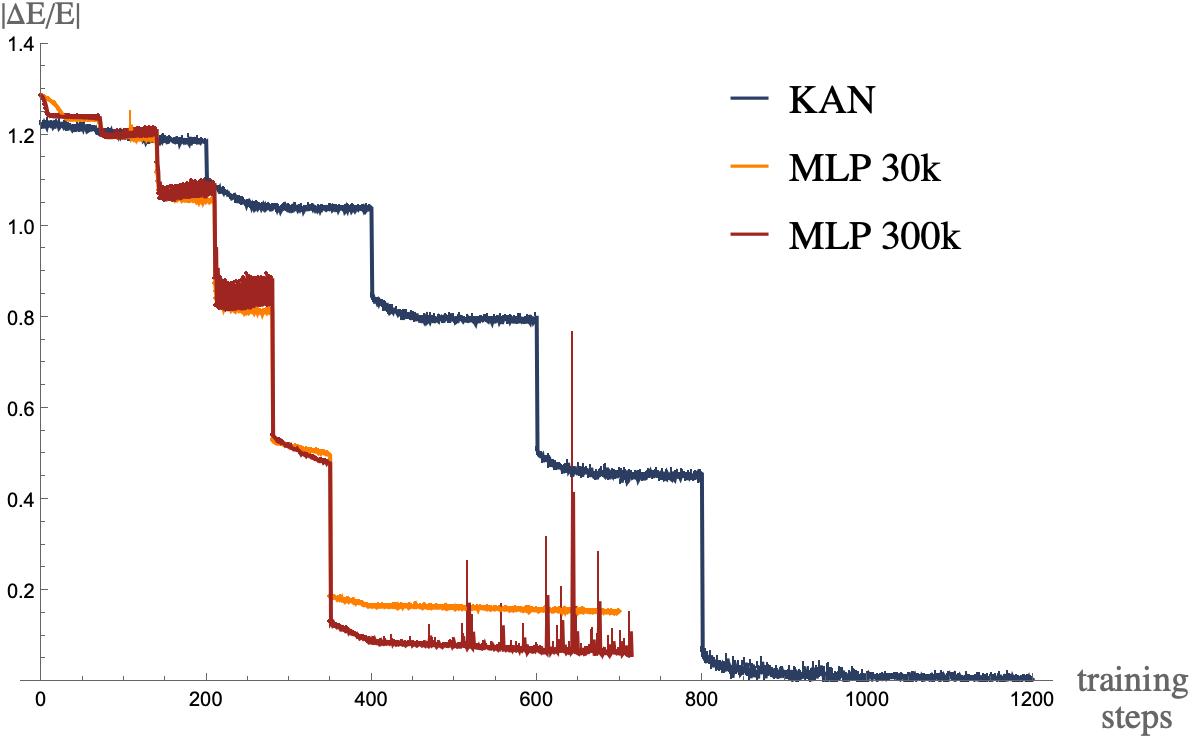} 
    \caption{Typical training process for both KAN and MLP networks. The coupling $g$ increases by steps from $g=0.1$ to the final value $g=0.5$ while the number of samples is also increased (from 2048 to 32768 in the KAN and from 64,000 to 1,000,000 in the MLP case). In addition, the number of knots in the splines in the KAN increases from 10 to 160. {\it After} the training is done it is easier to spot further improvements that could have been made to speed up the process. }
    \label{fig:KANvsMLP}
\end{figure}

Some experimentation is required to optimize the ramp up in $g$, spline resolution and  number of samples in a given problem but the numbers mentioned above are typical. In fact, {\it after} a training process is completed, one can usually identify improvements to speed it up. What we describe in this paper is the result of experimentation with a few particular models that would require adjustments in other contexts.

\begin{figure}
    \centering
    \includegraphics[width=0.6\linewidth]{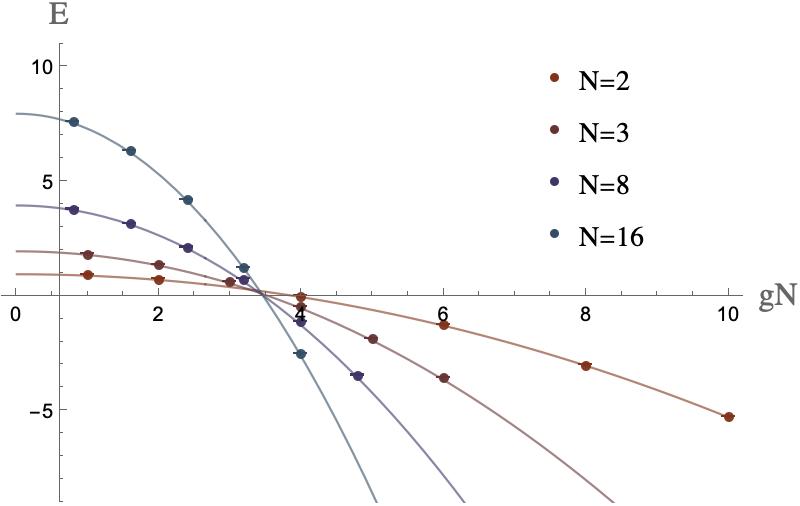}
    \caption{Energy for  model in \eq{eq:astra} with $g=-2\sigma$ for $N=2,4,8$ and $16$ particles plotted  against $gN$. The data points is the result of our calculation, the solid line shows the exact result of \eq{eq:astra-exact}.}
    \label{fig:2}
\end{figure}

A typical example of a training process is shown in  the blue curve of \fig{fig:KANvsMLP}. The steps where $g$, number of knots and samples, are visible. It is also visible that some speed up could be achieved by changing all these parameters sooner by a factor of ~2. Higher statistical precision in the last phase of training would also be useful. Typically, the training proceeds quickly until the last few percent of precision. At that point a substantial increase of statistics and number of knots, with a  corresponding increase in computer resources, is necessary. We will identify the source of this bottleneck and how to bypass it below. The correctness of the method and its implementation was validated by comparing our results to the exact one in \eq{eq:astra-exact} and the results are shown in \fig{fig:2}.

It is clear that KAN-based ansatz requires far fewer parameters than a MLP-based ansatz with similar accuracy. By itself, this is not a determining feature. Fewer parameters means fewer components of the energy gradient to compute but this is just one of many factors contributing to the overall computational cost. The energy landscape in parameter space is very different for the two ansatz and one might be much more amenable to training. Considering a MLP and a KAN-based ansatz providing similar accuracy, we find that the computational cost, measured in floating point operations per second (FLOPs), for one training step with 10,000 samples is roughly a factor 10 more costly for the MLP than for the KAN-based ansatz. This factor of 10 is roughly reflected on the walltime required for training, although this statement is highly dependent on details of the software and hardware implementation. The bottom line seems to be that, at least for the models/parameters we explored, the KAN-based ansatz is 10 times less costly.

\section{Improving short-distance behavior}
\label{sec:2body}

\begin{figure}[h]
    \centering
    \includegraphics[width=0.5\linewidth]{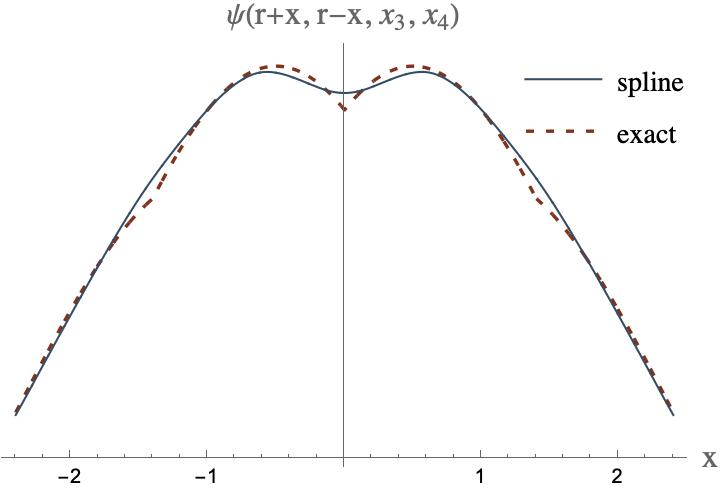}
    \caption{Exact and KAN wave-functions for $N=4, \ g=0.5, \ r=0.3, \ x_3=1, \ x_4=-1$, illustrating the difficulty of the splines to approximate the correct result near cusps at $x=0,\  x=(r-x_3)$, and $ x=(r-x_4)$. The vertical scale is arbitrary and shifted.}
    \label{fig:cusp}
\end{figure}

Due to the $\delta$-function interaction in \eq{eq:astra}, the exact wavefunction has a cusp when two particles occupy the same point in space. KAN's and splines can, given  enough number of knots, approximate this behavior arbitrarily well. For a a finite number of parameters, however, this behavior is difficult to capture. This problem is pervasive and not simply an artifact of our  unrealistic $\delta$-function potential.  
In any system with strong short-range potentials the derivative of the wavefunction changes quickly when two-particles are near each other, a behavior hard to capture numerically.
At short distances, in order to get a finite energy the divergent behavior of the potential is compensated by the kinetic energy contribution. This is called the Kato's condition \cite{Kato:1957bes}. An example is given in \fig{fig:cusp}. We verified that the poor behavior of our wavefunctions near the cusps is the main source of inaccuracy in the value of the energy. In the model in \eq{eq:astra} the contribution of the cusp region is the largest at intermediate values of $g$. At small $g$ the cusp (discontinuity of the derivative) is small while at large $g$, the cusp is pronounced but the value of the wavefunction at the cusp is small on account of the strong repulsion between particles.

We bypass this problem by adding to our ansatz an extra function $\kappa_2$ that is a function of the {\it difference} between the coordinates of each {\it pair} of particles:

\beq\label{eq:ansatz-2}
\kappa(x_1,...x_N) = \kappa_1(x_1,...,x_N) + \sum_{i<j} \nu(\tanh(c(x_i - x_j))) = \kappa_1(x_1,..x_N) + \kappa_2(x_1,...x_N),
\eeq where $\nu(x)$ is a new spline with additional variational parameters and $c(\Delta x)$ is a function we will specify below.
Obviously, the addition of the extra KAN can only increase the representability of the model, which remains universal in the limit of an infinite number of knots. The reason we believe it can be more efficient follows from the factorization property of the wavefunction: $\psi(x_1,x_2,\cdots,x_N)\xrightarrow[x_1\rightarrow x_2]{}
\phi(x_1-x_2) \Phi(\frac{x_1+x_2}{2}, x_3,\cdots,x_N)$. 
 In the particular example of the solvable model in \eq{eq:astra}, this factorization is satisfied exactly every time the distance between a pair of particles is smaller than the distance between their center of mass and each of the other particles. In general, this property is valid only on the $x_1\rightarrow x_2$ limit. The  function $\phi(x_1-x_2)$ is determined by two-body physics only and can be computed by solving the two-body Schr\"odinger equation (numerically or analytically) at zero energy\footnote{We will pursue elsewhere the idea of improving on this ansatz by adding a weak, analytically energy dependence on $\phi(x_1, \cdots, x_N)$ in the spirit of effective field theory.}.

\begin{figure}
    \centering    \includegraphics[width=0.7\linewidth]{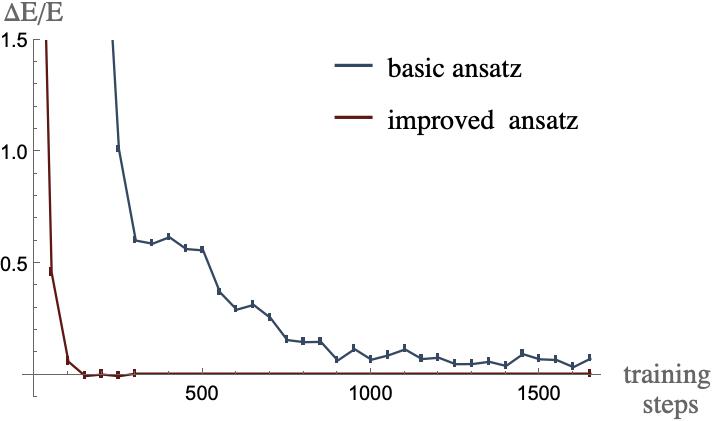}
    \caption{Training process for the model in \eq{eq:astra} with $N=32$, $g=0.1$. The blue  line was obtained with the ansatz in \eq{eq:kan-ansatz} and the red line 
  with the improved  ansatz in 
  \eq{eq:ansatz-2}. 
  Around step $250$ the improved ansatz gives the correct result up to stochastic uncertainties and the training was stopped.
  The values of the energy were computed  with a high statistics Monte Carlo but the intermediate values (not shown) and gradients were computed with lower statistical accuracy. }
    \label{fig:training}
\end{figure}

The function $c(\Delta x)$ can be subsumed into the spline $\nu(x)$ but, in order to make the spline as simple as possible to train we choose:
\beq
c(\Delta x) = \frac{mg|\Delta x|}
{1 + (mg\Delta x)^4}.
\eeq The shape of the function $c(\Delta x)$ is largely arbitrary; all that it is required is that it has the correct qualitative behavior of the small $\Delta x$ limit and vanishes at larger distances. We stress that the addition of the $\kappa_2(x_1,\cdots,x_N)$ ``cusp" term does not increase or decrease the family of functions included in the ansatz in the limit of infinite-resolution splines. This term, however, induces the wavefunction to satisfy Kato's condition \cite{kalos} with very smooth splines. For low resolution splines the $\kappa_2(x_1,\cdots,x_N)$ term is far superior in reproducing the correct ground state.
The ansatz in \eq{eq:ansatz-2} generates wavefunctions with the proper cusps.  This poses a challenge to Monte Carlo sampling, just like the $\delta$-function potential since most of the kinetic energy is concentrated in a few particle configurations with overlapping particles. This issue is easily bypassed by writing the kinetic term as $\braket{(\nabla^2\psi)/\psi} = -\braket{(\nabla\psi)^2/\psi^2}$ using integration by parts (the brackets stand for averaging over the $~|\psi|^2$ distribution).

The exact factorization property mentioned above makes the model in \eq{eq:astra} somewhat trivial: its exact ground state is entirely determined by the large distance and short distance behaviors already ``baked in" in the ansatz. This is demonstrated on \fig{fig:training} that shows how our improved ansatz is quickly trained to the exact result (within the statistical noise introduced by the Monte Carlo). For this reason, we consider now the following model instead:
\beq\label{eq:delta}
H = \sum_i^N \left(
-\frac{1}{2m} \frac{\partial^2}{\partial x_i^2} + \frac{1}{2}m\omega^2x_i^2
\right) + 
g \sum_{i<j}  \delta(x_i - x_j)\eeq This model has an analytic solution for the $N=2$ case \cite{busch} and some direct diagonalization results are available for the $N=3$ case \cite{harshman,damico}. The same model has also been studied  numerically  by several authors (for instance, \cite{doerte,doerte2, Kocik, Brouzos, Wilson}). The exact energy is known for the non-interacting case $g=0$ and in the $g\rightarrow\infty$ limit (Tonks-Girardeau limit), where the impenetrability of the particles creates nodes on the wavefunction just like a fermionic wavefunction \cite{tonks,girardeau}. This makes the intermediate $g$ regime the most interesting/challenging for numerical calculations.

\begin{figure}
    \centering
    \includegraphics[width=0.7\linewidth]{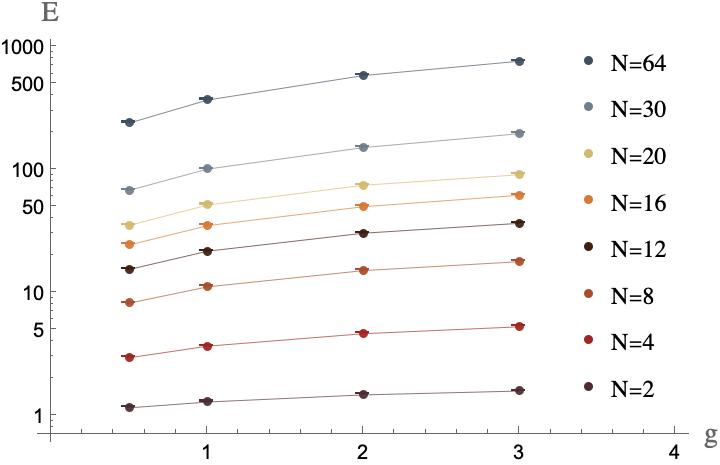}
    \caption{Ground state energy of the $\delta$-function plus harmonic trap model of \eq{eq:delta} as a function of coupling $g$ for different values of $N$. The error bars are, in most cases, too small to see and the lines are just to guide the eye.}
    \label{fig:Nxg}
\end{figure}


We arbitrarily set our accuracy goal around $1\%$, which can be comfortably achieved using laptops and small workstations. The uncertainties due to statistical noise in the final evaluation of the energy is much smaller than that and is shown as the error bars in the plots. 
We validated our methods, including our ability to recognize when training had achieved the $1\%$ accuracy level by comparing our calculations against the results of the aforementioned literature for $N=2, \cdots,6$. Our results for the ground state energy of the model \eq{eq:delta} over a range of couplings and particle numbers are summarized in \fig{fig:Nxg}. 
The limiting factor in accuracy does not stem from  the expressivity of the KAN ansatz but the cost of high precision Monte Carlo calculations.

We expect the two-body function $\nu$ to be determined by the short-distance behavior of the 2-particle Schr\"odinger equation and, therefore, dependent on the two-body interactions (the value of $g$) but not on the total number of particles. We verified that this is indeed true to high accuracy. In fact, not only the function $\nu$ of the the $N=2$ case be used as a starting point in the training of $\nu$ in the $N=64$ case but $\nu$ does not  have to be trained at all, its value extracted from a $N=2$ calculation being sufficient. This procedure is a practical method to implement the intuition that the many-body wavefunction is the result of many identical two-body collisions.

\section{Conclusions}\label{sec:conclusion}
In this paper we investigated the use of Kolmogorov-Arnold Networks  as wavefunction ansatz in variational quantum Monte Carlo (VMC) simulations for many-body quantum systems. 
We validated our numerical results  against exact solutions and prior literature, showing robust performance across varying particle numbers and coupling strengths. The study highlights the potential of KANs in efficiently simulating quantum systems with short-range forces, common in nuclear and atomic physics.
We compared their performance to traditional feed-forward neural networks (MLPs), demonstrating that KANs performance is more than competitive with MLPs.
While is impossible to define comparison criteria that are objective and useful, the evidence shows that KAN-based ansatz are roughly 10 times more efficient than MLP-based ones.

We introduced a modification to the KAN ansatz in order to represent cusps -- or any sudden variation of the wavefunction -- in an efficient way.  This modification of the ansatz to accommodate the cusps may seem {\it ad hoc} and particular to the models we considered, but we point out that the factorization property indicates otherwise and the same procedure should be useful every time strong, short-ranged potentials exist, including in higher dimensions. It also provides a method for transfer learning, that is, the two-particle function $\nu$ determined by solving the two-body problem can be used to speed up the simulation of systems with many particles. Avoiding the repetitive solution of the short-distance two-body problem within a many-body calculation is perhaps the greatest advantage of the present method, and we plan to extend it and systematize it in future work. In fact, it should be possible to use the same idea to incorporate $3, 4, \cdots$ particle short-distance correlations to simplify the many-body calculation.

\section{Acknowledgments}
This work was supported in part by the U.S. Department of Energy, Office of Nuclear Physics under Award Number DE-FG02-93ER40762. S.P. was partially supported by  DOE Grant KA2401045.

\section{Appendix: Variational Monte Carlo and short-range potentials}\label{sec:deltafunction}

We will here briefly review the Variational Monte Carlo (VMC) method and, in particular, the tricks required for a reliable calculation involving $\delta$-function potentials.

As in any variational calculation, one starts by considering a family of wavefunctions $\psi_\theta(x_1, \cdots, x_N)$ parameterized by $\theta$ ($\theta$ may stand for the weights and biases of a multi-layer perceptron or the spline parameters defining a KAN).
The energy and its gradient in relation to the parameters can be computed by the Monte Carlo method using: 

\bea
\mathcal{E}_\theta &=& \frac{\int dx_1...dx_N \; \psi^*_\theta H \psi_\theta}{\int dx_1...dx_N\; \psi^*_\theta\psi_\theta}
 = \frac{\int dx_1...dx_N \ |\psi_\theta|^2 \frac{H \psi_\theta}{\psi_\theta}}{\int dx_1...dx_N \psi_\theta^2}
 = \left\langle \frac{H \psi_\theta}{\psi_\theta} \right\rangle_{|\psi|^2} , \\
\frac{\partial \mathcal{E}}{\partial \theta} 
&=&
-2 \left\langle \frac{\partial \log(\psi_\theta)}{\partial \theta} 
\left(
\mathcal{E}_\theta-\frac{H\psi_\theta}{\psi_\theta} 
\right)\right\rangle_{|\psi|^2} , 
 \eea
here $\langle . \rangle_{|\psi|^2}$ denotes the average with respect to the probability $\psi^2$. It is important to note that the local gradient computation per sample depends on the global energy estimate and the quantity acts as a control variate and disappears as the $\psi_\theta$ approaches the true value \cite{Bhattacharya:2023pxx,Bedaque:2023ovz}.
In the presence of very short-ranged potentials the Monte Carlo evaluation of these integrals is difficult due to an overlap problem. Both integrands have support on a very small region while the Boltzmann factor $|\psi_\theta|^2$ is much more spread out. As a consequence, very few samples lie on the support of the integrand and the estimate of the integral is dominated by very few of the samples. This problem becomes critical for the $\delta$-function potential whose support has zero measure.

In \cite{neuralnetworkpaper} a solution for this problem was proposed. It is based on the re-writing of the integrals above as 
\bea
\langle \delta(x_1 - x_2)  \rangle_{|\psi|^2} 
&=& \frac{\int dx_1dx_2\cdots dx_N |\psi(x_1,x_2, \cdots, x_N)|^2 \delta(x_1 - x_2)}{\int dx_1dx_2\cdots dx_N |\psi(x_1,x_2,\cdots,x_N)|^2} \nn\\
&=&
\frac{\int dx_1 dx_3\cdots dx_N\; |\psi(x_1,x_1, x_3,\cdots, dx_N)|^2}{\int dx_1dx_2\cdots dx_N |\psi(x_1,x_2, \cdots,x_N)|^2} \nn\\
&=& \frac{\int dx_1dx_2\cdots dx_N |\psi (x_1,x_1,x_3,\cdots,x_N)|^2 D(x_2)}{\int dx_1dx_2\cdots dx_N |\psi(x_1,x_2,\cdots,x_N)|^2} \nn\\
&=& 
\frac{\int dx_1dx_2\cdots dx_N |\psi (x_1,x_1,x_3,\cdots,x_N)|^2 D(x_2) \frac{|\psi(x_1,x_2,\cdots,x_N)|^2}{|\psi(x_1,x_2,\cdots,x_N)|^2}}{\int dx_1dx_2\cdots,x_N |\psi(x_1,x_2,\cdots dx_N)|^2} \nn\\
&=& 
  \left\langle \frac{|\psi(x_1,x_1,x_3,\cdots,x_N)|^2}{|\psi(x_1,x_2,\cdots,x_N)|^2} D(x_2) \right\rangle_{|\psi|^2},
  \eea
where $D(x)$ is any function  such that $\int dx D(x) = 1$. We picked $D(x)$ to be a Gaussian with a width set by the length scale of the harmonic trap. The choice of width is very forgiving and the method works with any reasonable choice of $D(x)$.

\bibliographystyle{apsrev4-2}
\bibliography{biblio.bib}

\end{document}

%% file: mylatex.tex
\usepackage{graphics,setspace,epsfig,color}
\usepackage[letterpaper, dvips,width=7.5in,height=8.5in,includemp=false]{geometry}

\usepackage[vcentermath]{}
\usepackage{epsf}
\usepackage{amscd}
\usepackage{amsmath}
\usepackage{amsfonts}
\usepackage{amssymb}
\usepackage{braket}
\usepackage{bbm}

\usepackage[dvipsnames]{xcolor} 

\usepackage{tcolorbox}

\newcommand{\beq}{\begin{equation}}
\newcommand{\eeq}{\end{equation}}
\newcommand{\bea}{\begin{eqnarray}}
\newcommand{\eea}{\end{eqnarray}}

\newcommand{\nn}{\nonumber}

\graphicspath{{figures/}}
\def\beqamss#1\eeqs{\beq\begin{split} #1 \end{split}\eeq}

\newcommand{\eq}[1]{Eq.~\ref{#1}}
\newcommand{\fig}[1]{Fig.~\ref{#1}}

\newcommand{\sect}[1]{section~\ref{#1}}

\usepackage[colorlinks=true,backref=true, linktocpage=true,
citecolor=blue,urlcolor=blue,linkcolor=blue,pdfpagemode=UseOutlines]{hyperref}

\hypersetup{%
  bookmarksnumbered=true,
  pdftitle = {},
  pdfsubject = {},
  pdfauthor = {},
  pdfkeywords = {}
}